\documentclass[]{spie}
\usepackage[]{graphicx}

\title{Telescope to Observe Planetary Systems (TOPS): a high throughput 1.2-m visible telescope with a small inner working angle}

\author{Olivier Guyon\supit{a,b}, James R.P. Angel\supit{b}, Charles Bowers\supit{c}, James Burge\supit{b}, Adam Burrows\supit{b}, Johanan Codona\supit{b}, Thomas Greene\supit{d}, Masanori Iye\supit{e}, James Kasting\supit{f}, Hubert Martin\supit{b}, Donald W. McCarthy, Jr.\supit{b}, Victoria Meadows\supit{g}, Michael Meyer\supit{b}, Eugene A. Pluzhnik\supit{a}, Norman Sleep\supit{h}, Tony Spears\supit{i}, Motohide Tamura\supit{e}, Domenick Tenerelli\supit{j}, Robert Vanderbei\supit{k}, Bruce Woodgate\supit{c}, Robert A. Woodruff\supit{j}, Neville J. Woolf\supit{b}
\skiplinehalf
\supit{a}Subaru Telescope, NAOJ, 650 N. A'ohoku Pl., Hilo, HI, 96720, USA; \\
\supit{b}Steward Observatory, The University of Arizona, 933 N. Cherry Ave., Tucson, AZ  87521, USA\\
\supit{c}Goddard Space Flight Center, Greenbelt, MD  20771, USA\\
\supit{d}Ames Research Center, Moffett Field, CA  94035, USA\\
\supit{e}National Astronomical Observatory of Japan, Osawa 2-21-1, Mitaka, Tokyo 181-8588, Japan\\
\supit{f}Pennsylvania State University, Dept of Astronomy \& Astrophysics, 525 Davey Laboratory, University Park, PA  16802, USA\\
\supit{g}IPAC, Caltech, 770 S. Wilson, Pasadena, CA 91125, USA\\
\supit{h}Stanford University, Mitchell Building, room 373A, Stanford, CA 94305-2215, USA\\
\supit{i}LOGYX, 125 Kittoe Drive, Mountain View, CA 94043, USA\\
\supit{j}Lockheed Martin Space Corporation, P.O. Box 179, Denver, CO 80201-0179, USA\\
\supit{k}Princeton Univ., ORFE, Ace-42, E-Quad, Princeton, NJ  08544, USA}

\authorinfo{Further author information: (Send correspondence to Olivier Guyon.)\\Olivier Guyon.: E-mail: guyon@naoj.org, Telephone: 1 808 934 5901}

\begin{document} 
\maketitle 

\begin{abstract}
The Telescope to Observe Planetary Systems (TOPS) is a proposed space mission to image in the visible (0.4-0.9 $\mu$m) planetary systems of nearby stars simultaneously in 16 spectral bands (resolution R$\approx$20). For the $\approx$10 most favorable stars, it will have the sensitivity to discover $2 R_E$ rocky planets within habitable zones and characterize their surfaces or atmospheres through spectrophotometry. Many more massive planets and debris discs will be imaged and characterized for the first time. With a 1.2m visible telescope, the proposed mission achieves its power by exploiting the most efficient and robust coronagraphic and wavefront control techniques. The Phase-Induced Amplitude Apodization (PIAA) coronagraph used by TOPS allows planet detection at 2$\lambda/d$ with nearly 100\% throughput and preserves the telescope angular resolution. An efficient focal plane wavefront sensing scheme accurately measures wavefront aberrations which are fed back to the telescope active primary mirror. Fine wavefront control is also performed independently in each of 4 spectral channels, resulting in a system that is robust to wavefront chromaticity.  
\end{abstract}

\section{INTRODUCTION}
\label{sect:intro} 
There are now about 200 known exoplanets, most of them identified by indirect detection techniques. While detection of Earth-like planets is still out of reach for current instruments, it appears almost certain that such planets exist, and may even be in habitable zones of nearby stars \cite{kast93}. To both improve planet detection capabilities and allow characterization of their atmospheres and surfaces, we propose in this paper a visible 1.2-m diameter imaging telescope: the Telescope to Observe Planetary Systems (TOPS). TOPS would, for the first time, be able to directly image planetary systems similar to ours. It would identify and characterize massive rocky planets, gas giants and exozodiacal clouds around nearby stars.

To overcome the enormous star/planet contrast challenge, TOPS utilizes the most efficient coronagraphic and wavefront control techniques recently developed. The coronagraph adopted is the highly efficient Phase-Induced Amplitude Apodization (PIAA) coronagraph, described in \S\ref{sec:sls}. Wavefront control, described in \S\ref{sec:wfc}, relies upon focal plane wavefront sensing and a 2-step wavefront correction (primary mirror and fine DM). The spacecraft, described in \S\ref{sec:spacecraft}, are designed to meet the challenging wavefront stability requirements of this mission. Expected performance for both detection and characterization of exoplanets is quantified in \S\ref{sec:performance}.

\section{STARLIGHT SUPPRESSION}
\label{sec:sls}

\subsection{CORONAGRAPH CHOICES}
\begin{figure}
   \begin{center}
   \begin{tabular}{c}
   \includegraphics[height=7cm]{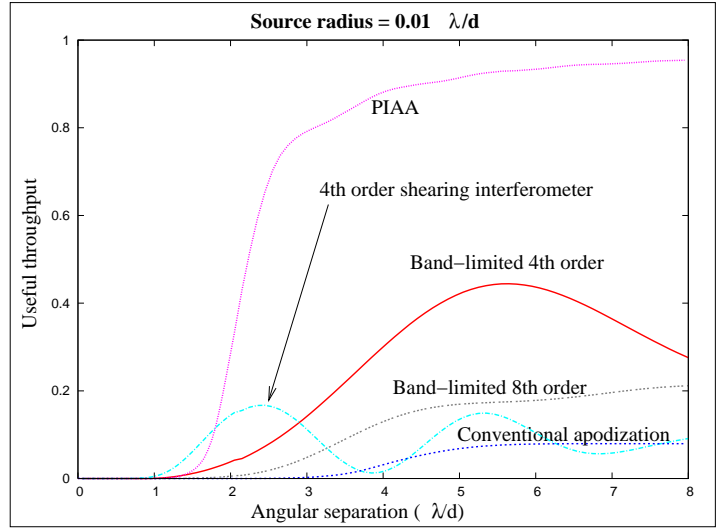}
   \end{tabular}
   \end{center}
   \caption[usefulT2] 
   { \label{fig:usefulT2} Radially averaged throughput of several coronagraphs at the 1e10 contrast level, for a 0.01 $\lambda/d$ stellar radius. Both the PIAA's throughput and its inner working angle are significantly better than those of coronagraphs derived from the Lyot design (Band-limited coronagraphs) or utilizing conventional pupil apodization.}
\end{figure}

The coronagraph must simultaneously achieve a high level of starlight suppression (typical star-to-planet flux ratios range from $10^9$ to $10^{11}$) and transmit most of the planet's light. Coronagraph performance is therefore primarily measured by the amount of detectable planet light spatially separated from the much stronger starlight in the image. 
This "useful throughput", radially averaged, is shown in Figure \ref{fig:usefulT2} as a function of angular separation for PIAA and three other coronagraphs which were considered for the Terrestrial Planet Finder Coronagraph (TPF-C) mission for imaging at the $10^{10}$ contrast level. This monochromatic simulation includes the effect of stellar angular size on the coronagraph performance.

This analysis illustrates the large range of performance between coronagraph designs: throughputs vary from $\approx$8\% (conventional pupil apodization) to nearly 100\%, and a PIAA coronagraph achieves the same throughput at 1.5 $\lambda/d$ as a conventional apodizer at 4 $\lambda/d$.
This figure shows that the PIAA approach, adopted for TOPS, is significantly more efficient than most other coronagraphs. It is the only currently known small-IWA coronagraph to offer sufficient robustness to stellar angular size to observe Earthlike systems with a 1.2m visible telescope. Its development truly constitutes a major leap forward in coronagraphy, and is currently the most practical option to approach the "theoretical" performance limit imposed by diffraction.

\subsection{THE PIAA CORONAGRAPH}
\begin{figure}
   \begin{center}
   \begin{tabular}{c}
   \includegraphics[height=8cm]{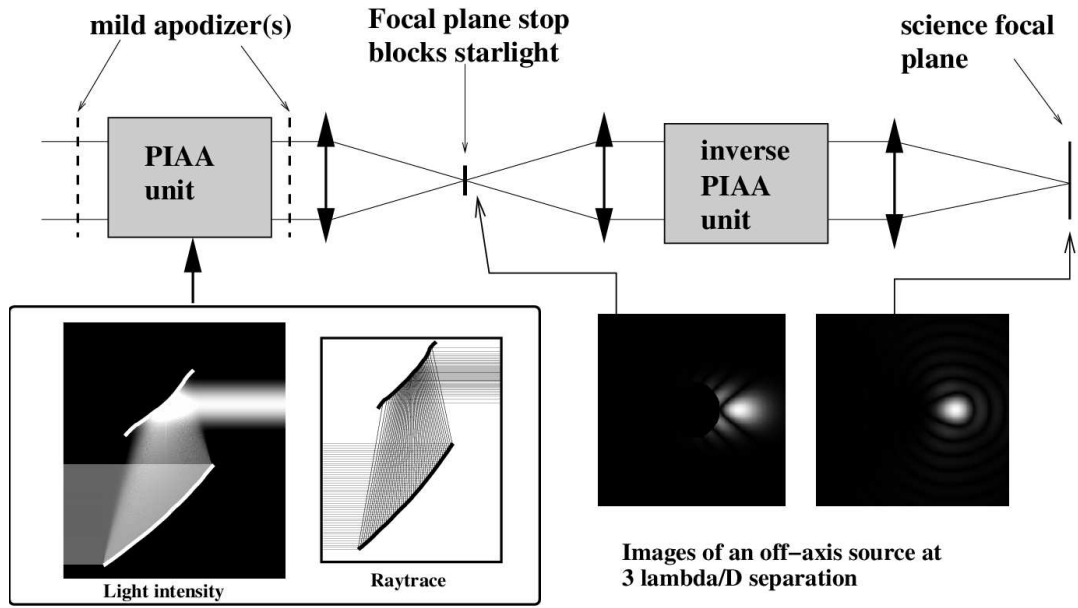}
   \end{tabular}
   \end{center}
   \caption[example] 
   { \label{fig:piaa_principle} 
Schematic representation of the PIAAC. The telescope light beam enters from the left and is first apodized by the PIAA unit. Mild apodizer(s) are used to perform a small part of the apodizations, and are essential to both mitigate chromatic diffraction propagation effects and allow for the design of ``friendly'' aspheric PIAA mirrors. An high contrast image is then formed, allowing starlight to be removed by a small occulter. An inverse PIAA unit is required to ``sharpen'' the image of off-axis sources.}
   \end{figure}

PIAA coronagraphy \cite{guyo03,traub03,guyo05a,vand05,mart06,vand06,pluz06} utilizes geometric reflection on aspheric mirrors to perform a lossless apodization of the telescope beam (see Figure \ref{fig:piaa_principle}). This apodization is designed to deliver $10^{10}$ PSF contrast at and beyond 1.5$\lambda/d$ in the coronagraphic focal plane, where stellar light is blocked by an occulter. After removal of the bright starlight, the original beam geometry is restored by "inverse-PIAA" optics to deliver aberration-free diffraction-limited planet images over a sufficiently wide field of view.

The main characteristics of this coronagraph are:
\begin{enumerate}
\item{Achieves high contrast ($10^{10}$ for a perfect wavefront)  at small IWA (from 1.5$\lambda/d$ to 2$\lambda/d$, depending on the exact design adopted). The ability to search for planets at this angular separation (0.15" at the blue end of TOPS spectral coverage) greatly improves the number of giant planets we expect to image and is essential to access the habitable zone of nearby stars.} 
\item{Nearly 100\% throughput and 360-degree search area. The low intrinsic luminosity of planets imposes strong requirements on the throughput and efficiency of the coronagraph. At and beyond 2 $\lambda/d$ (slightly more for a star of large angular diameter), the PIAAC transmits more than 2/3 of the planetary photons, regardless of position angle.}
\item{Robustness to stellar angular size. Unlike other small-IWA ($< 2 \lambda/d$) coronagraphs, the PIAAC performance is unaffected by stellar angular size and small ($\approx 0.01 \lambda/d$ or less) pointing errors, provided that the focal plane occulter is slightly oversized ($\approx 1.8 \lambda/d$ in our baseline design).}
\item{Preserves the telescope's angular resolution. Thanks to the absence of an undersized Lyot pupil plane stop, the planet light is confined into a single $\lambda/d$ - wide diffraction core. This is critical to minimize the amount of zodiacal and exozodiacal light mixed with the planet light, and provides additional robustness against confusion with exozodiacal features, coronagraphic leaks, background sources and other planets.}
\item{Can be designed with high level of achromaticity. Our proposed "hybrid" PIAA design\cite{pluz06} relies on both PIAA apodization and mild conventional apodization. It can be designed to maintain high contrast simultaneously in a broad spectral band, at the cost of a small loss of light (about 10\%) in the mild apodizer.}
\end{enumerate}
In addition to points [1], [2] and [3], which have already been factored in Figure \ref{fig:usefulT2}, the superior angular resolution (point [4]) widens the performance gap between conventional coronagraphs and the PIAAC. When all factors are taken into account, {\bf the PIAAC offers the same performance level that conventional coronagraphs on a telescope $\approx$2 times smaller in diameter}. Examples of "conventional" coronagraphs include the $8^{th}$ order band-limited coronagraph \cite{kuch05} and conventional pupil apodization \cite{kasd03}, both considered for the TPF-C mission.

\subsection{CORONAGRAPH DESIGN FOR TOPS}
\label{ssec:hybrid}
A system which entirely relies upon PIAA to apodize the telescope beam requires very ``unfriendly'' aspheric mirror shapes, with a strong curvature at the edge of the first PIAA mirror. This results in optics serious manufacturability challenges and strong chromatic diffraction propagation effects which restrict the spectral bandwidth across which high contrast can be maintained. In a pure-PIAA system, the useful spectral bandwidth at the $10^{10}$ contrast is extremely small, and direct detection of ETPs is practically impossible\cite{vand06}.
Since this problem originates from the need to produce the very dark outer edge of the apodized beam, it can be efficiently mitigated by designing PIAA optics with do not fully apodize the beam\cite{pluz06} (typically leaving the edge of the apodized beam at 1\% of the surface brightness of the center of the apodized beam) and using conventional apodization to complete the apodization. The conventional apodizer(s) may be placed before and/or after the PIAA unit, and can be either continous or binary apodization mask(s).
The fraction of light absorbed by the conventional apodizer(s) is an important parameter in the design of the PIAAC, as it defines the chromaticity of the PSF contrast. With about 10\% of the light absorbed in the conventional apodizer(s), the PIAAC can be designed to deliver $10^{10}$ PSF contrast across reasonably wide spectral bands \cite{pluz06,pluz06a}. 

We adopt for TOPS this ``hybrid'' PIAA design. In this design, maintaining high contrast in a wider spectral bandwidth requires more light to be absorbed in the conventional apodizer. Rather than tuning the hybrid PIAA optics to deliver $10^{10}$ contrast across the full TOPS spectral coverage, the PIAA optics are designed to maintain $10^{10}$ contrast only in one of the 4 spectral channels. The conventional apodizer of the hybrid PIAA is moved after the dichroics, and can therefore be designed independently for each spectral channel. Similarly, each fine DM can independently correct for small wavefront phase differences between the channels. 
TOPS therefore utilizes 4 hybrid PIAA coronagraphs, each tuned for the corresponding spectral channel, which share a common set of 2 PIAA mirrors. This approach optimizes the tradeoff between throughput, spectral coverage and optical complexity, and reduces the risk of single point failures impacting the mission's science return.

\section{WAVEFRONT CONTROL}
\label{sec:wfc}

\subsection{REQUIREMENTS AND OVERALL ARCHITECTURE}
\begin{table}[h]
\caption{Wavefront control requirements for $10^{10}$ contrast. Wavefront tolerances are given at the entrance of the PIAA. A coronagraph system including PIAA, focal plane occulter and inverse PIAA was simulated at 550 nm to derive these requirements. The sampling time necessary to measure the corresponding level of aberration at SNR=5 is given here for a $m_V=5$ star.}

\label{tab:wfc_requ}
\begin{center}       
\begin{tabular}{|l|l|l|l|}
\hline
Mode & Required control accuracy & Sensor & SNR=5 sampling time \\
\hline
Tip / tilt & 0.9 nm rms/mode & LOWFS & 0.5 s \\
\hline
Focus & 43 pm rms & LOWFS & 1 s \\
\hline
Astigmatism & 70 pm rms/mode & LOWFS & 1 s \\
\hline
Mid spatial frequencies & 1.5 pm rms/mode & Science CCDs \& LOWFS & 5 min\\
                        & [$\approx$40 pm rms total, & &\\ 
                        & 15 pm per actuator] & &\\ 
\hline
High spatial frequencies & Strehl ratio $>0.98$ & none, relies on optical & - \\
                         &                      & quality of components &  \\
\hline
\end{tabular}
\end{center}
\end{table}

\begin{figure}
   \begin{center}
   \begin{tabular}{c}
   \includegraphics[height=18cm]{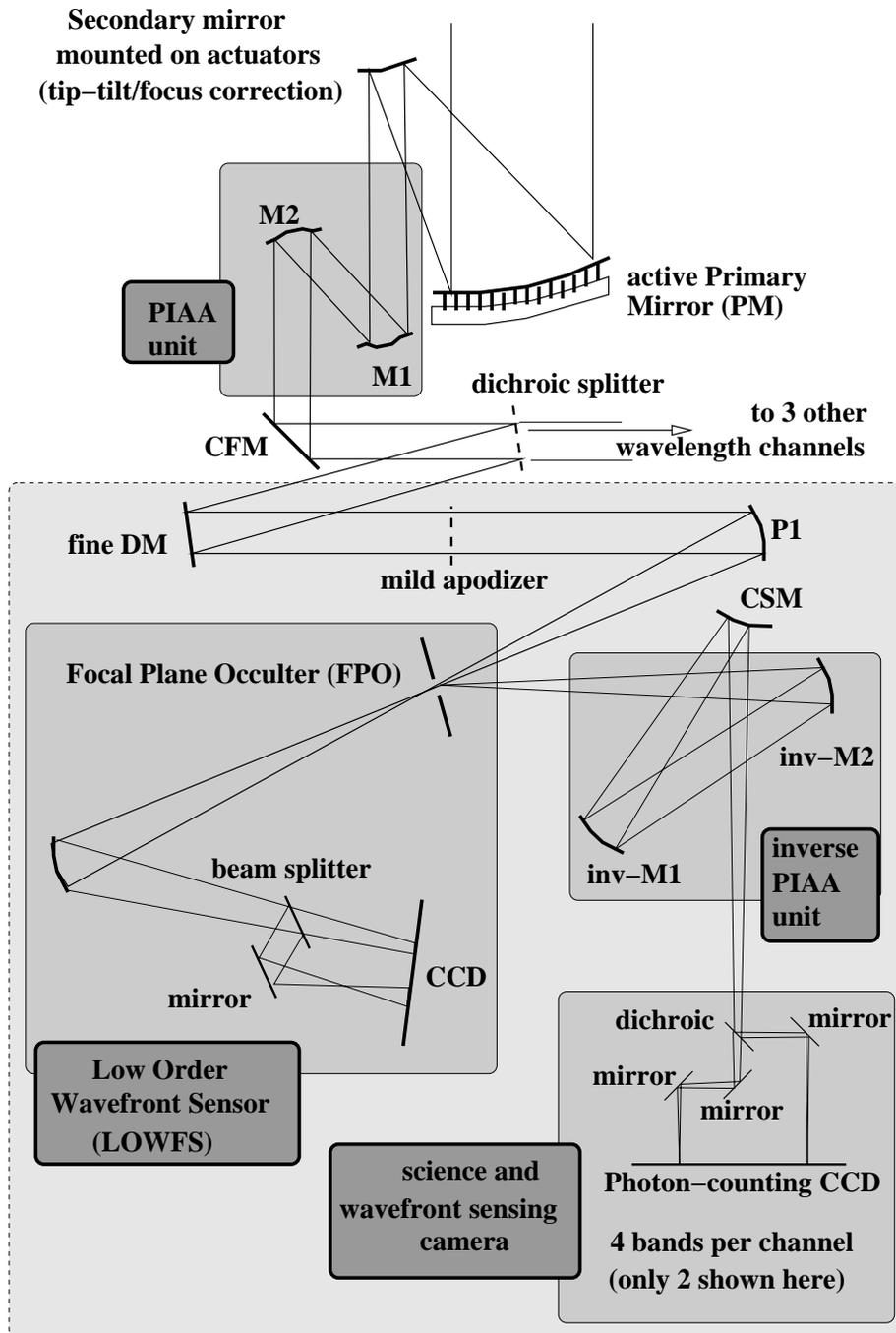}
   \end{tabular}
   \end{center}
   \caption[tops_principle] 
   { \label{fig:tops_principle} 
Schematic representation of the combined coronagraphic and wavefront control architecture for TOPS. The primary mirror (PM), secondary mirror (SM) and common fold mirror (CFM) are the only optics common to all 4 spectral channels. The light gray box (bottom half of the figure) shows one on the 4 spectral channels. A fine DM and mild apodizer allow independent phase and amplitude correction within each spectral channel. The focal plane occulter (FPO) divides light between the low order wavefront sensor (LOWFS), which uses the central $\approx 1.6 \lambda/d$ radius in the focal plane and the science camera. In the science camera path, an inverse PIAA unit removes the field aberrations introduced by the PIAA unit. Within each of the 4 spectral channels, the light is further divided in 4 spectral bands with dichroics and mirrors in front of the detector, resulting in 16 (4x4) spectral bands.}
   \end{figure}

Wavefront control requirements to maintain a $10^{10}$ contrast are given in Table \ref{tab:wfc_requ}. These requirements must be satisfied up to the focal plane occulter. The overall coronagraph and wavefront sensing architecture for TOPS is designed to meet these requirements, and is shown in Figure \ref{fig:tops_principle}. Wavefront correction (\S\ref{ssec:actelem}) is achieved in 2 successive steps: (1) the active primary mirror first corrects for large aberrations and (2) fine DMs within each spectral channel perform fine correction and correct for aberrations which are not common to all spectral channels. 

\subsubsection{Wavefront Stability}
Table \ref{tab:wfc_requ} is setting requirements on the wavefront stability for a $m_V$=5 star. These requirements drive the TOPS design in 2 directions:
\begin{itemize}
\item{The TOPS optics (particularly the primary mirror) must be intrinsically extremely stable and free of thermal distortions or vibrations. The choice of the orbit, pointing strategy, spacecraft design and the Disturbance-free payload, described in \S\ref{sec:spacecraft}, are all critical to provide a stable environment for the telescope optics.}
\item{Efficient wavefront sensing schemes need to be used to minimize the sampling time required to detect these aberrations. The coronagraph high throughput is extremely helpful: if a 10\% throughput coronagraph were used, it would take 10 times longer to measure the wavefront aberrations to the same accuracy.}
\end{itemize}
The most critical requirement is for mid-spatial frequencies, which should be held stable (or at least should be predictable) to 1.5pm per mode (15pm per actuator) within 5 minutes. For brighter stars, wavefront sensing is faster and stability requirements are relaxed.

\subsubsection{Wavefront Chromaticity}
Within each spectral channel (slightly more than 0.1 $\mu$m wide), the wavefront must be achromatic to levels given in Table \ref{tab:wfc_requ}. Excessive wavefront chromaticity could only allow high contrast on one subchannel (R$\approx$20) at a time per channel. TOPS addresses this issue by (1) performing wavefront correction directly on the primary mirror to minimize chromatic Fresnel propagation of wavefront aberrations (see \S\ref{sssec:pm}) and (2) providing independent wavefront control in 4 spectral channels (see \S\ref{ssec:hybrid} and \S\ref{sssec:finedm}).

\subsection{ACTIVE ELEMENTS}
\label{ssec:actelem}
\subsubsection{PRIMARY MIRROR (PM)}
\label{sssec:pm}
Conventional space optical systems designed for optical diffraction limited imaging require a wavefront accuracy of ~ 30 nm rms. The PM can have larger errors, if the system wavefront 
is corrected with a smaller mirror conjugated to the primary, as in the Hubble Space Telescope. However, for TOPS the wavelength-dependent intensity variations in the pupil caused by propagation from an aberrated PM cannot easily be corrected. The relay optics to the deformable mirror need to be extremely high precision to avoid propagating chromatic amplitude errors, and will likely result in restricted bandwidth for the extreme high contrast. Thus we apply active correction of errors by active control at the PM with performance and simplicity benefits.
Active control of PM figure is new for space but is standard for large ground based telescopes. All of the current generation of large telescopes ($>$ 6 m diameter) have active primaries with upwards of 100 actuators. In addition, adaptive secondary mirrors up to 1 m diameter with hundreds of actuators and millisecond response time are now coming into use for ground telescope adaptive optics. These mirrors have 2 mm thick, aspheric meniscus facesheet made at the University of Arizona (UA) Mirror Lab. 
   \begin{figure}
   \begin{center}
   \begin{tabular}{c}
   \includegraphics[height=7cm]{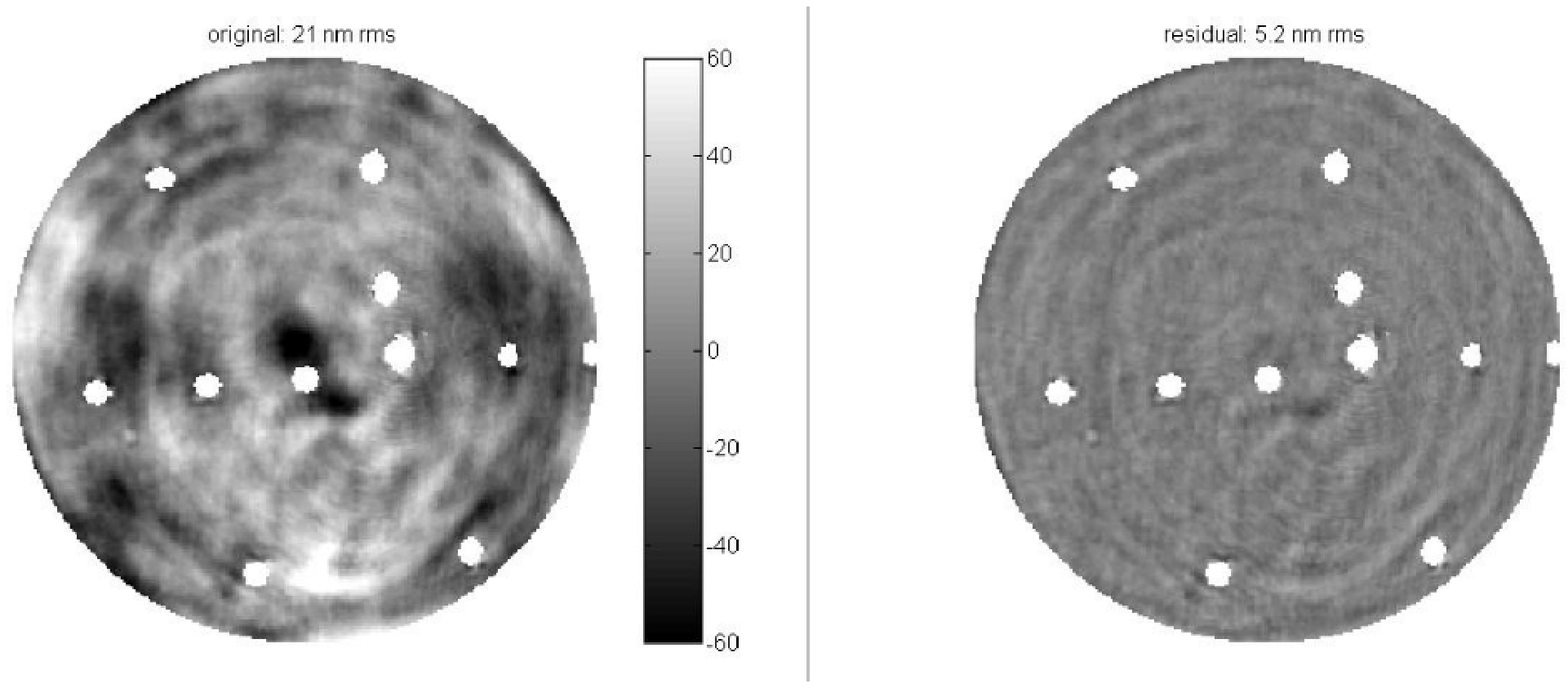}
   \end{tabular}
   \end{center}
   \caption[activepm] 
   { \label{fig:activepm} Central 1.2 m aperture of a 1.7 m off-axis mirror whose parent is a 5 m f/0.7 paraboloid. Left: measured figure near the completion of polishing with an actively stressed lap. Right: predicted figure after ion figuring with a 30 mm diameter ion beam. Several small defects in the measurement system have been masked out of the figures.}
   \end{figure}
For TOPS, a relatively small number of PM actuators, 111 in all, is sufficient, since the search region is only a few diffraction widths in radius.

Current optical finishing technology can produce a figure accuracy better than 5 nm rms for highly aspheric off-axis mirrors. Figure \ref{fig:activepm} shows an example of such a mirror that is being figured at the Steward Observatory Mirror Lab with a stressed-lap polishing system, and includes a simulation of the effect of ion figuring as a final step in the process. The stressed-lap system achieves an accuracy on the order of 20 nm rms rapidly on fast aspheres, while ion figuring is one of several technologies that give very predictable removal of small-scale errors with amplitudes of several hundred nm. With effort, residual errors can be reduced to the order of 1 nm rms, at which level they would likely be dominated by gravity deflections.

The adaptive primary mirror baselined for TOPS will use thermal actuators.  The mirror will be built and figured in just the same way as current space telescope primaries, structured as a lightweight honeycomb sandwich.  A surface accuracy of 5 nm rms can be expected, thus actuation of $\pm$ 20 nm should be sufficient to null out the errors.  Glass of small but finite expansion coefficient will be used, so that shape changes can be introduced with thermal gradients.  Even for a passive mirror of nominally zero coefficient glass, good to $10^{-8}$/K, active thermal control to ~ 1 mK would be required to avoid variable halo speckles.  In our case, we exploit the thermal sensitivity to control
the figure.  Temperatures will be set locally by balancing radiative cooling to a sink slightly below the operating temperature (eg with a cold finger into each honeycomb cell) with resistive heating on the ribs and facesheets. The expansion coefficient of the glass used to make the ribs and facesheets will be chosen separately for optimum control.  Thus for control of intermediate order modes, the 15 cm deep ribs will be made of fused silica of thermal expansion coefficient of 5x$10^{-7}$ .  A  4 nm local bump, for example, would thus require heating by 100mK.  At the same time, for low order control, the facesheets will be made of lower coefficient glass, for example titania doped silica with coefficient 5x$10^{-8}$ .  Then a uniform 100 mK temperature difference induced between the front a back facesheet temperatures would cause curvature of 6 nm sagittal depth across the 1.2 m primary.  Thus by active thermal controls of order 100 mK controlled to 0.1 mK the figure errors will be reduced from a few nm to a few pm.

\subsubsection{CORRECTION BY FINE DEFORMABLE MIRRORS}
\label{sssec:finedm}
A 32x32 actuator DM, located in each spectral channel after its dichroic filter, further improves the PM correction and extends the correction to higher spatial frequencies.  This dichroic DM also corrects those small aberrations that cannot be addressed by the PM, including:
\begin{itemize}
\item{Chromatic aberrations due to diffraction propagation between PM, SM and the PIAA optics (these aberrations are expected to differ between spectral channels at mid-spatial frequencies)}
\item{Aberrations not common to all wavelength channels, introduced at and after the dichroics}
\end{itemize}
Although the roles of PM and the fine dichroic DMs are mostly complementary, one of them can to some degree make up for sub-optimal performance of the other. The fine DMs are placed after the PIAA unit rather than before to create a well defined "dark hole" at the coronagraph's focal plane occulter. Wavefront control to create such a ``dark hole'' has been successfully demonstrated in the High Contrast Imaging Testbed at JPL\cite{trau03}.

The fine DMs are also an integral part of the wavefront sensing scheme (see \S\ref{ssec:wfc_sensing}).

\subsection{WAVEFRONT SENSING}
\label{ssec:wfc_sensing}
\subsubsection{FOCAL PLANE WAVEFRONT SENSING}
Wavefront sensing is also performed by the science focal plane arrays. In this scheme, akin to phase diversity, known offsets are sent to the DMs upstream of the coronagraph occulter. The corresponding variations in the post-coronagraphic image speckles provide the signal to reconstruct the pre-coronagraph wavefront, or, equivalently, the complex amplitude of the focal plane speckle field. The science focal plane array is immune to non-common path errors and aliasing effects, thus yielding wavefront sensing exactly where diffracted light needs to be canceled. This enables picometer accuracy wavefront sensing of mid-spatial frequencies in closed loop. The wavefront reconstruction algorithm maximizes the likelihood that the acquired data corresponds to a given wavefront (the wavefront is the "free" parameter).

To achieve the $10^{10}$ contrast in the science image, the optics must be stable to the levels given in Table \ref{tab:wfc_requ} within the amount of time necessary for wavefront sensing. Mid-spatial frequencies are especially critical, and must be stable to better than 1.5 pm per mode (40 pm overall RMS) within a few minutes.
The TOPS focal plane science detectors are photon-counting CCDs. The temporal wavefront correction bandwidth is therefore set by photon noise: a few minutes at the $10^{10}$ contrast level for a $m_V=5$ target. Information is combined between all spectral channels to maximize sensitivity to "fast" (few min or faster) wavefront changes expected to originate from the primary mirror (the only large optical element in the system) and flexure-induced misalignments. Differential aberrations between spectral channels are reconstructed at a much slower rate as they are expected to be largely static and the number of photons available per channel is small. Predictive control algorithms will efficiently correct for slow drifts in the wavefront (caused by variable thermal gradients).

\subsubsection{LOW ORDER WAVEFRONT SENSING AND FOCAL PLANE MASK}
   \begin{figure}
   \begin{center}
   \begin{tabular}{c}
   \includegraphics[height=7cm]{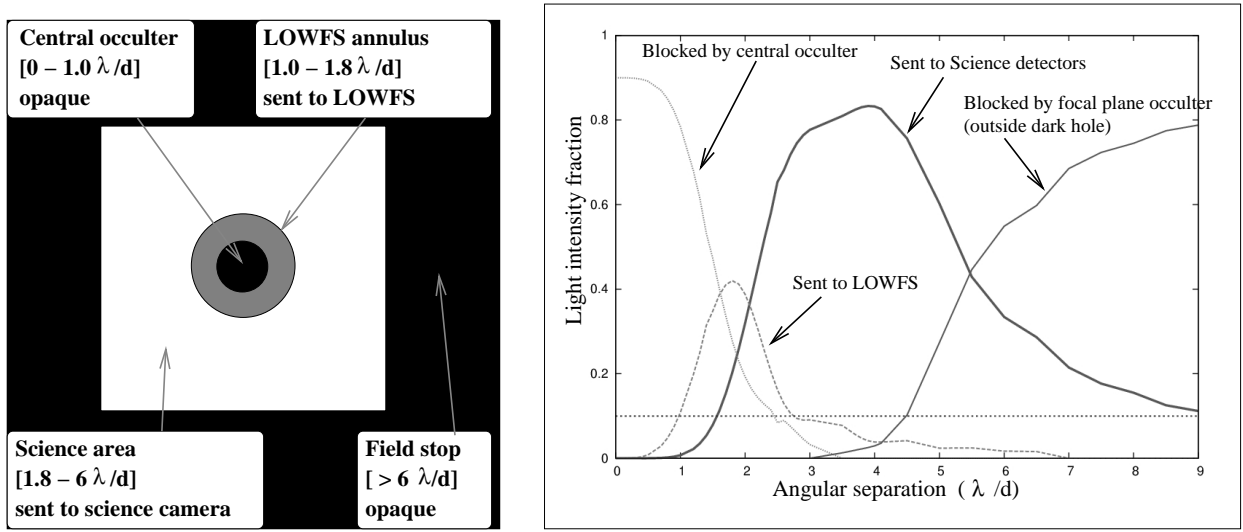}
   \end{tabular}
   \end{center}
   \caption[transm] 
   { \label{fig:transm} The focal plane mask (left) distributes the light between the LOWFS and the science path. Fraction of the light which is blocked by the central occulter, sent to the LOWFS, sent to the science path and blocked by the field mask as a function of angular separation for a point source.}
   \end{figure}

\begin{figure}
   \begin{center}
   \begin{tabular}{c}
   \includegraphics[height=6cm]{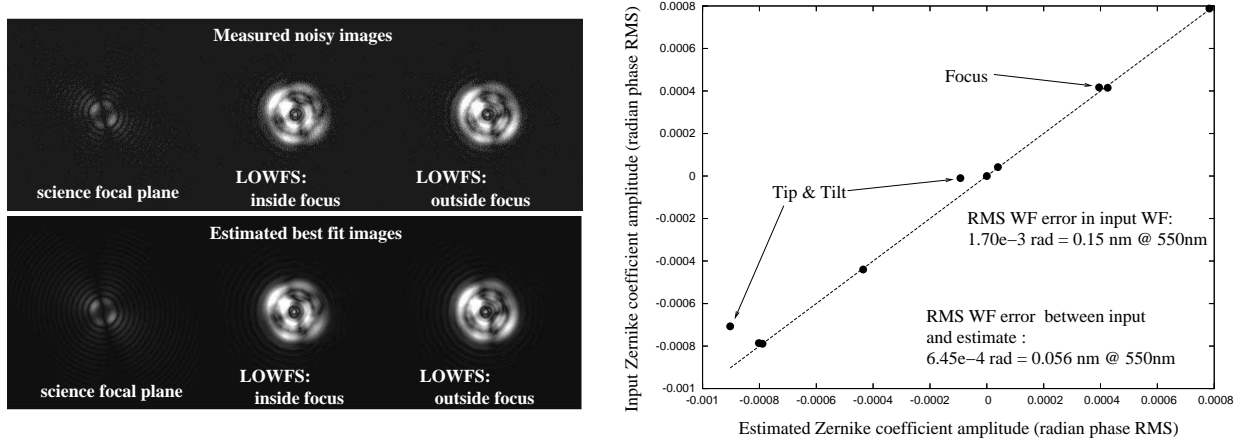}
   \end{tabular}
   \end{center}
   \caption[lowfs] 
   { \label{fig:lowfs} Noisy inside and outside focus images obtained by the LOWFS (top left). Photon noise is included in this simulation, where $10^{12}$ photons entered the telescope pupil. These images were simulated at 550nm with a 50pm error per Zernike for the first 10 Zernikes. Wavefront reconstruction from these 2 frames accurately recovers low order wavefront aberrations (right), to an accuracy level which exceeds the requirement for $10^{10}$ contrast. Simulated science focal plane and LOWFS (bottom left), computed from the wavefront estimation, match the noisy images acquired by the LOWFS (top left).}
   \end{figure}

For each channel, light extracted by the focal plane occulter (Figure \ref{fig:transm}) is re-imaged to accurately measure low-order aberrations, such as tip-tilt and focus. Optimal sensitivity is obtained when the very center of the occulter is blocked: only the $\approx$ 1 to 1.8 $\lambda/d$ annulus around the optical axis is re-imaged. Inside and outside focus images of this annulus are obtained simultaneously by the beam splitter and mirror arrangement, as shown in Figure \ref{fig:lowfs}.

The ability of the LOWFS to produce large relative intensity variations for small wavefront errors is key to its ability to measure low order aberrations. The LOWFS does not rely on precise calibration of optical components or camera. This is illustrated in Figure \ref{fig:lowfs}, where 50-pm aberration per low-order mode are clearly visible without processing. Wavefront reconstruction modeled with this data shows a level of correction beyond the TOPS requirements. The LOWFS allows sensing, and therefore correcting, low-order aberrations before they start to "leak" into the focal plane science detectors.

The same focal plane mask also serves as a field stop, blocking light which is outside the ``dark hole'' created by the fine DM. Without this field stop, bright speckles outside the dark hole would be radially smeared by the inverse-PIAA unit and would interfere with planet detection.

\section{SPACECRAFT}
\label{sec:spacecraft}

\subsection{OVERALL OPTICAL DESIGN AND PACKAGING}

   \begin{figure}
   \begin{center}
   \begin{tabular}{c}
   \includegraphics[height=7cm]{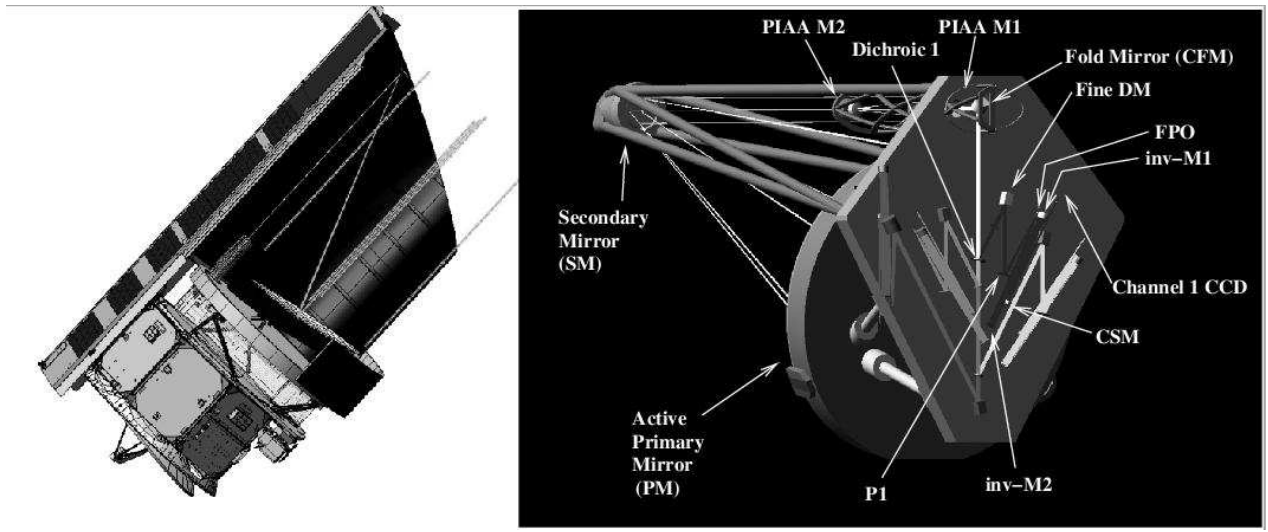}
   \end{tabular}
   \end{center}
   \caption[tconnor] 
   { \label{fig:tconnor} TOPS spacecraft (left) and details of the optics layout (right).}
   \end{figure}
   
As shown in Figure \ref{fig:tconnor}, an afocal OTA followed by an afocal PIAA efficiently applies apodization while converting the beam to component compatible sizes. Thus the OTA reduces the beam from 1200 mm only to 100mm diameter for the PIAA M1, allowing us to optimize its size for fabrication. The PIAA M2 mirror is easier to fabricate, so the beam size is reduced to match the fine DM (from 10 to 30mm depending from DM technology choice). The Common Fold Mirror (CMF) directs the resultant beam to the dichroic filter chain.
Three dichroic filters in the collimated beam create four spectral channels. A DM follows its channel's dichroic filter to correct the accumulated wavefront error in its channel from the PM through its dichroic filter(s). In each channel, a single parabolic mirror P1 images the corrected field at a reflective FPO of ~1.6 $\lambda_{mid}$/D radius. The FPO mechanism co-aligns the FPOs by translating them precisely in lateral plane. The FPO reflects the beam to inverse PIAA mirrors, followed by the common science mirror (CSM) that image the field at the channel-dedicated CCD with pixel sampling $\lambda_{mid}$/(3d).  A passive bandpass defining unit lies directly in front of the CCD in each channel to form spectral bands, eliminating the need for four filter wheels and maintaining high overall throughput. A single passive radiative cooler on the space side of the facility cools the eight CCDs to $\approx$ 150 K operation temperature by radiation to space to reduce their dark current to $\approx$ 0.4 e-/pix/hr to support zero read noise photon counting.

Fig. \ref{fig:tconnor} shows the spectral splitting configuration using near normal incidence (15 degrees) dichroic filters to minimize polarization effects, maintain high efficiency across the channel, and reduce spectral loss in the dichroic's spectral transition region. Efficient coatings, large detector QE, and efficient apodization with the PIAA, result in a very efficient system.


\subsection{VIBRATION CONTROL}
The spacecraft is designed to provide a stable environment for the TOPS optics, which is essential for successful wavefront control at the sub-nm level. Vibrations at frequencies above $\approx 0.1Hz$ would be especially detrimental to TOPS performance, as they could not be accurately sensed and corrected by the wavefront control system (limited by photon noise).

A key design feature of the TOPS spacecraft is the Disturbance Free Payload (DFP). In this architecture\cite{pedr03}, the Optical Telescope Assembly (OTA) and spacecraft are separate bodies: the OTA flies within the spacecraft, and the two bodies interact through non-contact sensors and actuators. The OTA is therefore isolated from spacecraft vibrations (reaction wheels) down to zero frequency, and pointing control and stability is significantly improved. High level of vibration isolation with the DFP concept has been successfully demonstrated in laboratory \cite{pedr02,gonz04,pedr05}.

\section{SCIENCE GOALS, OBSERVING PROGRAM AND EXPECTED PERFORMANCE}
\label{sec:performance}
The TOPS science goals are:
\begin{itemize}
\item{Determine the distribution of extrasolar giant planets in the solar neighborhood and characterize their physical properties beyond what is possible with radial velocity and transit techniques}
\item{Determine planetary system architectures from the brightness, structure and spectral energy distribution of low surface density zodiacal dust clouds around nearby stars}
\item{Search and characterize rocky planets in the habitable zone of nearby stars comparable to our own}
\end{itemize}

Simulations for mission performance are used to identify targets with the best chance of success (highest probability of planet detection/characterization in a given observation time) for both rocky planets and Gas Giants. It is assumed here that the TOPS photometric efficiency is 9\% (0.85 optics throughput x 0.85 detector QE x 0.5 due to wavefront sensing overhead x 0.5 due to polarization overhead x 0.5 field of view overhead), that detection requires the planet's maximum elongation to reach 2 $\lambda/d$, and that residual starlight calibration cannot recover planets at contrast higher than $10^{11}$. Table \ref{tab:targetnumb} shows for different type of planets how many targets are potentially available as a function of exposure time.

\begin{table}[h]
\caption{Number of theoretically accessible targets (R=5,SNR=5) for different planet types. The ``HZ'' is used here as a unit of distance: it is proportional to the square root of the stellar bolometric luminosity and equal to 1 AU for our Sun.}
\label{tab:targetnumb}
\begin{center}       
\begin{tabular}{|l|l|l|l|}
\hline
Planet type & 40 mn & 4 hr & 4 day\\
\hline
Earth @ 1 HZ & 2 & 2 & 4 \\
\hline
Earth @ 1.8 HZ & 2 & 2 & 8 \\
\hline
10 Earth mass @ 1.8 HZ & 2 & 5 & 31 (10 for 1 day)\\
\hline
10 Earth mass @ 1 HZ & 3 & 3 & 9 \\
\hline
Jupiter @ 1 AU & 10 & 10 & 10\\
\hline
Jupiter @ 5 AU & 69 & 183 & 482 \\
\hline
Jupiter @ 1 HZ & 10 & 10 & 10 \\
\hline 
\end{tabular}
\end{center}
\end{table}

\begin{figure}
   \begin{center}
   \begin{tabular}{c}
   \includegraphics[height=6cm]{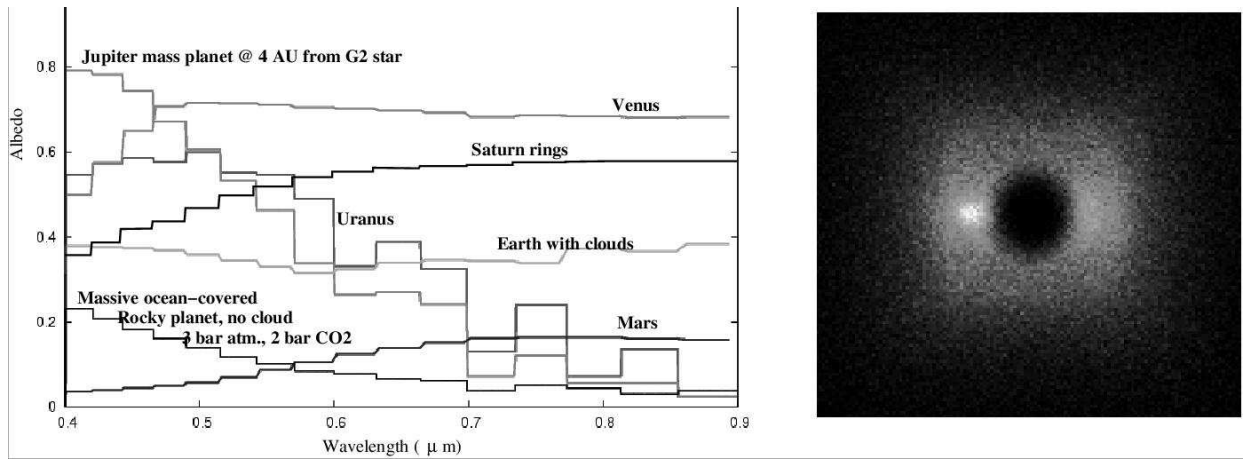}
   \end{tabular}
   \end{center}
   \caption[allspectra] 
   { \label{fig:allspectra} Left: Spectra of several planetary system bodies binned to the TOPS resolution; the TOPS R$\approx$20 resolution allows identification of the planet atmosphere and/or surface. Right: Simulated TOPS visible (0.5 - 0.6 $\mu$m) image of a 1.7 Earth mass planet at 1.2 AU from tau Ceti (3.65pc). A zodiacal cloud at 1.5x solar system level is also visible in this 8hr exposure.}
   \end{figure}

The simulation tool used to generate Table \ref{tab:targetnumb} includes all major effects: phase functions, zodiacal background and exozodiacal light cloud (here assumed to be at a similar level as our local zodiacal cloud). 
These results, all derived in the 0.5-0.6 micron band, show that massive rocky planets could potentially be detected in the habitable zone of up to 30 stars (assuming 1 day per observation), while gas giants at 5 AU separation could be identified around several hundred targets. These numbers have been used to outline a possible observing plan for the baseline mission:
\begin{itemize}
\item{{\bf Deep survey [300 days]}: The target list for this survey consists of ~30 stars. For at least 10 of these (Table \ref{tab:targetnumb}), we can detect within a 1 day exposure a large rocky planet (10 Earth masses) at the outer edge of the habitable zone (1.8 HZ) - see Figure \ref{fig:allspectra}, right. Allowing 1 day on average per exposure, and 10 visits per target to determine orbital parameters, this deep survey would require approximately 300 days (1 year) of observing time without overheads.}
\item{{\bf Shallow survey [76 days]}:
The target list for this survey consists of 183 stars, the ones for which a Jupiter at 5 AU can theoretically be detected within a 4 hour exposure. Some of these stars are included in the Deep survey, and therefore do not require additional observation. Allowing 1 hours on average per exposure (corresponding to a minimum SNR=5 detection in the 0.5-0.6 micron band for the best $\approx$80 targets), and 10 visits per target, 1830 hours (76 days) would be required for this survey. }
\item{{\bf Planets characterization \& monitoring [function of number of planets detected and mission duration]}:
Higher signal-to-noise ratio observations will be acquired for selected targets. As illustrated in Figure \ref{fig:allspectra}, left, $R\approx 20$ spectrophotometry allows characterization of the planet atmosphere and/or surface. TOPS will be able to image and characterize several already known Radial Velocity planets.}
\end{itemize}

\acknowledgments     
Concept development for TOPS has been supported at Arizona by NASA under Terrestrial Planet Finder Instrument Concept Study grant NNG05GO57G. Development of the PIAA coronagraph and focal plane wavefront control techniques is supported by the National Astronomical Observatory of Japan and JPL contract numbers 1254445 and 1257767 for Development of Technologies for the Terrestrial Planet Finder Mission.



\begin{thebibliography}{}
\bibitem[1]{kast93} Kasting, J.F., Whitmire, D.P., Reynolds, R.T.  1993, Icarus, 101, 108
\bibitem[2]{kasd03} Kasdin, N.J., Vanderbei, R.J., Spergel, D.N., Littman, M.G.  2003, ApJ, 582, 1147
\bibitem[3]{guyo03} Guyon, O.  2003, A\&A, 404, 379
\bibitem[4]{traub03} Traub, W.A., Vanderbei, R.J., 2003, ApJ, 599, 695
\bibitem[5]{guyo05a} Guyon, O., Pluzhnik, E.A., Galicher, R., Martinache, F., Ridgway, S.T., Woodruff, R.A.  2005, ApJ, 622, 744
\bibitem[6]{vand05} Vanderbei, R.J., Traub, W.A.  2005, ApJ, 626, 1079
\bibitem[7]{mart06} Martinache, F., Guyon, O., Pluzhnik, E.A., Galicher, R., Ridgway, S.T.  2006, ApJ, 639, 1129
\bibitem[8]{vand06} Vanderbei, R.J.  2006, ApJ, 636, 528
\bibitem[9]{pluz06} Pluzhnik, E.A., Guyon, O., Ridgway, S.T., Martinache, F., Woodruff, R.A., Blain, C., Galicher, R. 2006, ApJ, in press, astro-ph/0512421
\bibitem[10]{kuch05} Kuchner, M.J., Crepp, J., Ge, J.  2005, ApJ, 628, 466
\bibitem[11]{pluz06a} Pluzhnik, E.A., Guyon, O., Warren, M., Woodruff, R.A., Ridgway, S.T.  2006, SPIE 6265-135  
\bibitem[12]{trau03} Trauger, J.T., Moody, D., Gordon, B., Gursel, Y., Ealey, M.A., \& Bagwell, R.B.  2003, Proc. SPIE, 4854, 1
\bibitem[13]{pedr03} Pedreiro, N.  2003, Journal of Guidance, Control and Dynamics, 26, 794
\bibitem[14]{pedr02} Pedreiro, N., Carrier, A.C., Lorell, K.R., Roth, D.E.  2002, American Institute of Aeronautics and Astronautics, 2002-5027
\bibitem[15]{gonz04} Gonzales, M.A., Pedreiro, N., Roth, D.E., Brookes, K., Foster, B.W.  2004,  American Institute of Aeronautics and Astronautics, 2004-5247
\bibitem[16]{pedr05} Pedreiro, N., Gonzales, M.A., Foster, B.W., Tankle, T.L., Roth, D.E.  2005, American Institute of Aeronautics and Astronautics, 2005-5876
\end{thebibliography}
\end{document}